# Double- and Single-Sided bis-Functionalization of Graphene

K. C. Knirsch,[†] R. A. Schäfer,[†] F. Hauke and A. Hirsch*

Kathrin C. Knirsch,[†] Ricarda A. Schäfer,[†] Dr. Frank Hauke, Prof. Dr. Andreas Hirsch
Department of Chemistry and Pharmacy & Joint Institute of Advanced Materials and Processes (ZMP)
University of Erlangen-Nürnberg
Henkestr. 42, 91054 Erlangen, Germany
E-mail: andreas.hirsch@fau.de

**Abstract:** For the first time bis-functionalization of graphene employing two successive reduction and covalent bond formation steps are reported. Both bulk functionalization in solution and functionalization of individual sheets deposited on surfaces have been carried out. Whereas in the former case attacks from both sides of the basal plane are possible and can lead to strain-free architectures, in the latter case, retro-functionalizations can get important when the corresponding anion of the addend represents a sufficiently good leaving group.

Multiple covalent functionalization of synthetic carbon allotropes with different addends represents an attractive concept for the design and combination of specific chemical, physical, and materials properties of fullerenes, carbon nanotubes, and graphene. As a consequence, also promising practical applications of nanocarbons such as sensors,[1] nanocomposites,[2] and biomedical products[3] can be targeted. The construction of molecular architectures consisting of a carbon allotrope core and two or more different covalently attached functionalities has been realized so far in fullerene[4-10] and carbon nanotube[11-15] chemistry, whereas examples for mixed graphene derivatives are still elusive. Among the most efficient methods for graphene functionalization is the reduction/exfoliation of graphite with alkaline metals in suitable solvents followed by quenching of the intermediately formed graphenides with electrophiles (Scheme 1a). We have recently shown that in this way alkylated, arylated, and hydrogenated graphenes with high degrees of addition at both sides of the basal plane can be prepared.[16-18] In the case of related reductive functionalizations of carbon nanotubes and fullerenes, we[19] and others[20] have demonstrated that such reactions can be reversible. It has not been shown yet but it is reasonable to assume that related reversible processes such as that depicted in Scheme 1b can also play a role in graphene chemistry. Also it is important to keep in mind that: a) covalent fullerene and carbon nanotube chemistry is monotopic (only exohedral addend binding can take place), whereas graphene chemistry can be both monotopic (when the sheets are supported on a surface) or ditopic (in homogeneous solution); b) exhaustive homotopic additions will eventually lead to an increase of strain energy (eclipsing addend interactions, deviation from normal bond angles); c) ditopic addition can lead to more stable and less strained geometries including complete strain-free graph**a**ne with an all chair conformation of the six-membered C-rings.[18, 21-22] It is to be expected that the degree of retro-functionalization (Scheme 1b) represents an interplay between the strain energy of the graphene adduct itself and the stability of the leaving group R⁻. We now present for the first time a) a bis-functionalization sequence of graphene and b) the investigation of topicity and leaving group dependence of the retro-functionalization. These investigations constitute a fundamental piece of carbon allotrope chemistry.

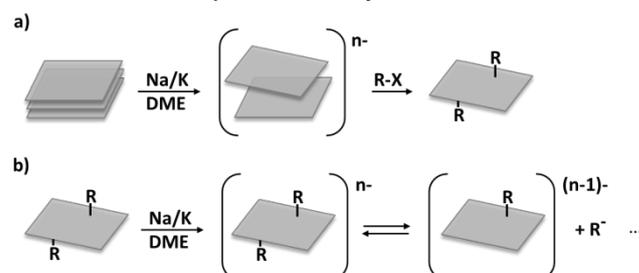

**Scheme 1.** a) Reductive functionalization and b) reductive retro-functionalization of graphene.

As subsequent addition reactions we choose the treatment of graphenides with diazonium salts[16] and alkyl iodides.[17] In order to investigate the influence of graphene topicity, we carried out the addition to graphenides in homogeneous dispersion (double-sided bulk functionalization) and to CVD graphenide supported on a Si/SiO$_2$ surface (single-sided functionalization). For the bulk functionalization, pristine natural graphite **G$_P$** was exfoliated by wet chemical reduction using Na/K alloy in 1,2-dimethoxyethane (DME). After this activation the negatively charged graphenide sheets were treated with the first electrophile. After work-up a second activation with Na/K alloy was initiated followed by the addition of the second electrophile. For the synthesis of 4-methoxyphenyl-hexyl-graphene **G$_{AB}$**, we used 4-methoxyphenyldiazonium tetrafluoroborate **A** as the first and *n*-hexyl iodide **B** as the second electrophile (Scheme 2).

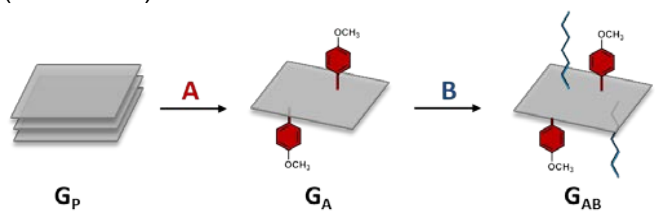

**Scheme 2.** Stepwise reductive activation of graphite **G$_P$** and **G$_A$** with Na/K alloy in DME and subsequent addition of the electrophiles 4-methoxyphenyldiazonium tetrafluoroborate **A** and *n*-hexyl iodide **B**, leading to the bis-functionalized graphene derivative **G$_{AB}$**.

Thermogravimetric analysis coupled with mass spectrometry (TGA-MS) (Figure 1) revealed a mass loss of -20.6 % (black curve) in **G$_{AB}$** compared with the starting material (grey curve). The large mass loss correlates with the characteristic fragments of the 4-methoxyphenyl- as well as hexyl-units detected by mass spectrometry. A peak for *m/z* 39, 77, 78, 107 and 108 (see Scheme S1) is observed at about 510 °C (see Figure S1, top and center). These mass fragments can be assigned to allyl- (*m/z* 39), phenyl- (*m/z* 77, 78) and methoxyphenyl-units (*m/z* 107, 108) which clearly demonstrates the successful binding of methoxyphenyl addends introduced by the corresponding diazonium precursor. The mass fragments *m/z* 43 (propyl), 57 (butyl), 71 (pentyl) and 85 (hexyl) (see Scheme S1) peaking at 480 °C (see Figure S1, bottom) on the other hand prove the alkylation by *n*-hexyl iodide.

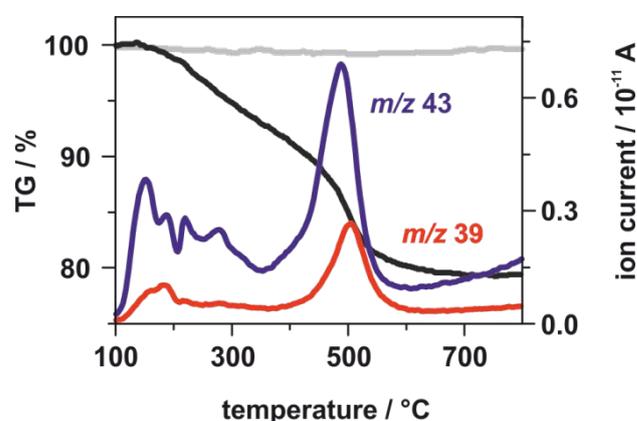

**Figure 1.** TG profile of the 4-methoxyphenyl-hexyl-functionalized reaction product **G$_{AB}$** (black) and pristine graphite **G$_P$** (grey) along with characteristic mass fragments of **G$_{AB}$** which can be assigned to the aryl- (*m/z* 39) and alkyl groups (*m/z* 43).

We then carried out statistical Raman spectroscopy (SRS)[23] as a very powerful tool to investigate the covalent addend binding and thus the composition of **G$_{AB}$**. In order to analyze the basal *sp$^3$*-C-centers formed by the covalent addend binding, we determined the mean ratio of the defect induced D-band intensity and the G-band intensity (mean $I_D/I_G$-ratio). This ratio serves as a measure for the degree of functionalization. A mean $I_D/I_G$-ratio of 0.7 (Figure 2a, Table S2) was observed for the first covalent adduct **G$_A$** (cf. **G$_P$** in Figure S2, Table S2). In the final mixed adduct **G$_{AB}$** the mean $I_D/I_G$-ratio increased further to 1.4 (Figure 2b, Table S2). At the same time the statistical distribution of the $I_D/I_G$-ratio (Figure 2b) experiences a considerable narrowing compared with that of **G$_A$** (Figure 2a). We also carried out the same reaction sequence but in reversed order. For this purpose graphite **G$_P$** was first transferred into alkylated graphene **G$_B$** and then arylated to give the mixed adduct **G$_{BA}$** (Scheme 3).

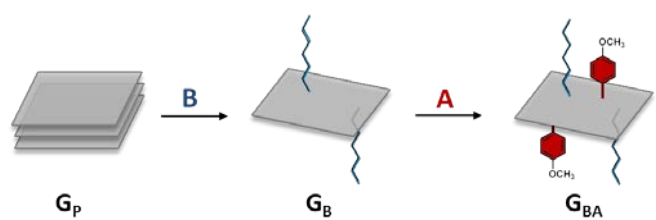

**Scheme 3.** Stepwise reductive activation of graphite **G$_P$** and **G$_B$** with Na/K alloy in DME and subsequent addition of the electrophiles *n*-hexyl iodide **B** and 4-methoxyphenyldiazonium tetrafluoroborate **A**, leading to the bis-functionalized graphene derivative **G$_{BA}$**.

Again a narrowing of the statistical distribution of the $I_D/I_G$-ratio was observed after the second functionalization which reveals a rather high homogeneity of the samples **$G_{AB}$** and **$G_{BA}$** (Figure 2). The intermediate **$G_B$** displayed a mean $I_D/I_G$-ratio of 1.6 (Figure 2c, Table S2), whereas the final product **$G_{BA}$** exhibited a mean $I_D/I_G$-ratio of 1.3 (Figure 2d, Table S2). In this context it is important to see that simultaneously to the decreasing mean $I_D/I_G$-ratio going from **$G_B$** to **$G_{BA}$**, a broadening of the Raman bands emerges. The same behavior is found for **$G_{AB}$**.

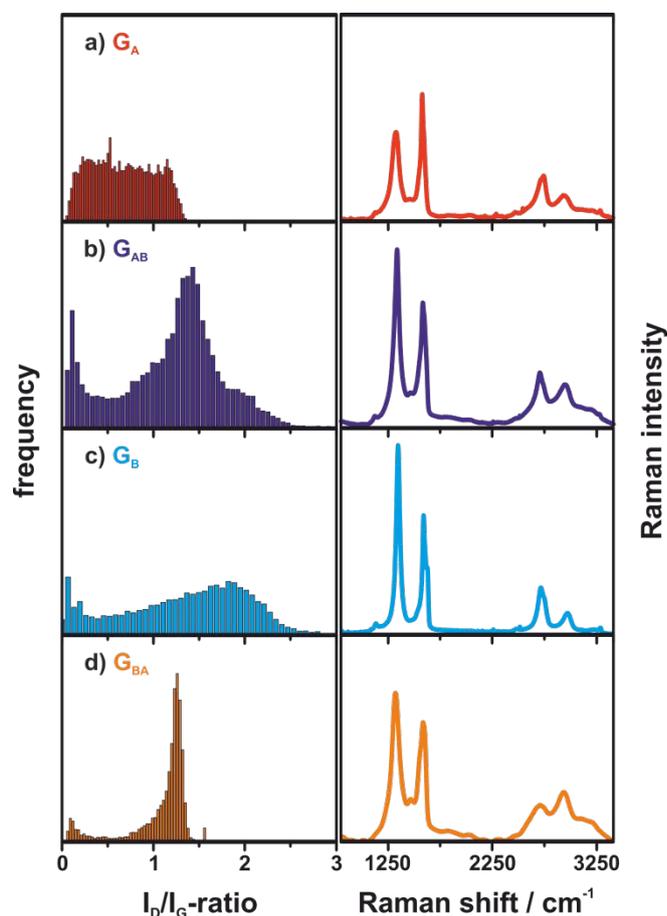

**Figure 2.** Left: Raman histograms ($I_D/I_G$-ratio) of **$G_A$**, **$G_{AB}$**, **$G_B$**, and **$G_{BA}$**. Right: averaged spectra of the respective samples, $\lambda_{exc.}$ = 532 nm.

In fundamental studies of Lucchese[24] and Cançado,[25] a general relationship between the $I_D/I_G$-ratio and the mean defect distance $L_D$ (Figure 3) was determined that exhibits a maximum $I_D/I_G$-ratio at a certain $L_{D\text{-crit}}$. The $I_D/I_G$-ratios can be assigned to regimes of either above or below $L_{D\text{-crit}}$. In order to determine the $L_D$ under consideration, the width of the Raman bands has to be analyzed. A broadening of the Raman bands especially in the region between 2,200-3,300 cm$^{-1}$ is known to stem from a very high defect density and a decreased mean defect distance $L_D$ in the basal graphene plane (Figure 3).[23, 25] In our cases a similar $L_D$ of the final products **$G_{AB}$** and **$G_{BA}$** which is below $L_{D\text{-crit}}$ was obtained, irrespective of the reaction sequence order.

These results also show that retro-functionalization processes such as that depicted in Scheme 1 play a minor role during the entire reaction sequence leading to the mixed products **$G_{AB}$** and **$G_{BA}$**. Functionalization is preferred over retro-functionalization. Obviously, the final products are thermodynamically rather stable since double-sided attacks can take place. For example, a 1,2-addition to a double bond in graphene taking place from opposite sides (see also Scheme 5a) leads to an almost strain-free addend binding.

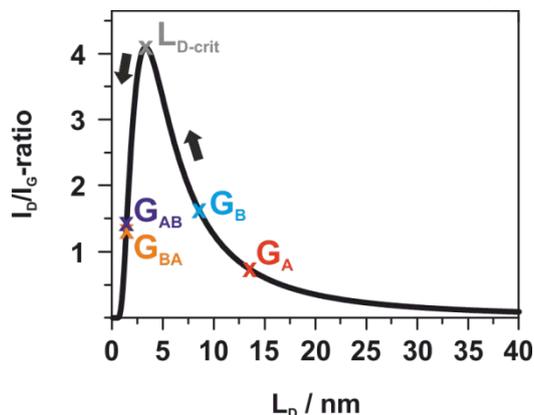

**Figure 3.** Correlation between the $I_D/I_G$-ratio and the mean defect distance $L_D$ of **G$_A$**, **G$_{AB}$**, **G$_B$**, and **G$_{BA}$**, $\lambda_{exc.}$ = 532 nm.[25]

In the next step we wanted to analyze the subsequent single-sided addition reactions to graphene. For this purpose we used CVD graphene deposited on a Si/SiO$_2$ substrate. Therefore, the pristine CVD graphene **G$_P$(CVD)** on Si/SiO$_2$ was reduced with a drop of the before mentioned Na/K-DME solution and afterwards a drop of the first electrophile in DME was added (Scheme 4). After a washing step with DME, *i*-prop, water, and acetone, the resulting functionalized graphene was activated again and the second electrophile was added.

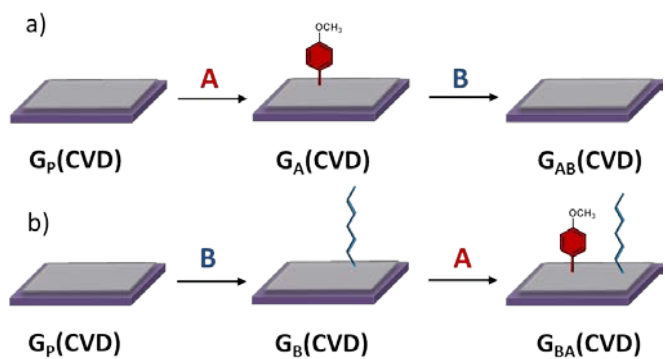

**Scheme 4.** a) Stepwise reductive activation of CVD graphene on a Si/SiO$_2$ substrate **Gp(CVD)** and **G$_A$(CVD)** with Na/K alloy in DME and subsequent addition of the electrophiles 4-methoxyphenyldiazonium tetrafluoroborate **A** and *n*-hexyl iodide **B**, leading to **G$_{AB}$(CVD)** and b) the reversed addition route leading to **G$_{BA}$(CVD)**.

SRS showed that pristine CVD graphene **G$_P$(CVD)** exhibits a mean I$_D$/I$_G$-ratio of 0.2 arising from domain boundaries (further Raman data see ESI, Figure S3-7, Table S3, 4).[26] The reduction itself did not lead to a change of the I$_D$/I$_G$-ratio (Figure S8). After the addition of *n*-hexyl iodide, however, the mean I$_D$/I$_G$-ratio increased from 0.2 (**G$_P$(CVD)**) to 0.6 (**G$_B$(CVD)**) (Figure 4c, S11; Table S3, 4). This behaviour demonstrated the successful alkylation leading to single-sided **G$_B$(CVD)**. The following arylation via the corresponding diazonium salt caused a further increase of the mean I$_D$/I$_G$-ratio to 0.9 (**G$_{BA}$(CVD)**) (Figure 4d, S12; Table S3, 4). Obviously, also the next step, namely the single-sided bis-functionalization takes place (Scheme 4). To investigate whether the bis-functionalization can also be accomplished via the reversed order, we first allowed the diazonium salt to react with the CVD graphenide **G$_P$(CVD)**. The intermediate **G$_A$(CVD)** exhibited a I$_D$/I$_G$-ratio of 1.0 (Figure 4a, S9; Table S3, 4). Significantly, after carrying out the second treatment (reductive activation, addition of the *n*-hexyl iodide), we observed a decrease of the mean I$_D$/I$_G$-ratio to 0.3 (Figure 4b, S10; Table S3, 4), being very close to the value of unfunctionalized **G$_P$(CVD)**.

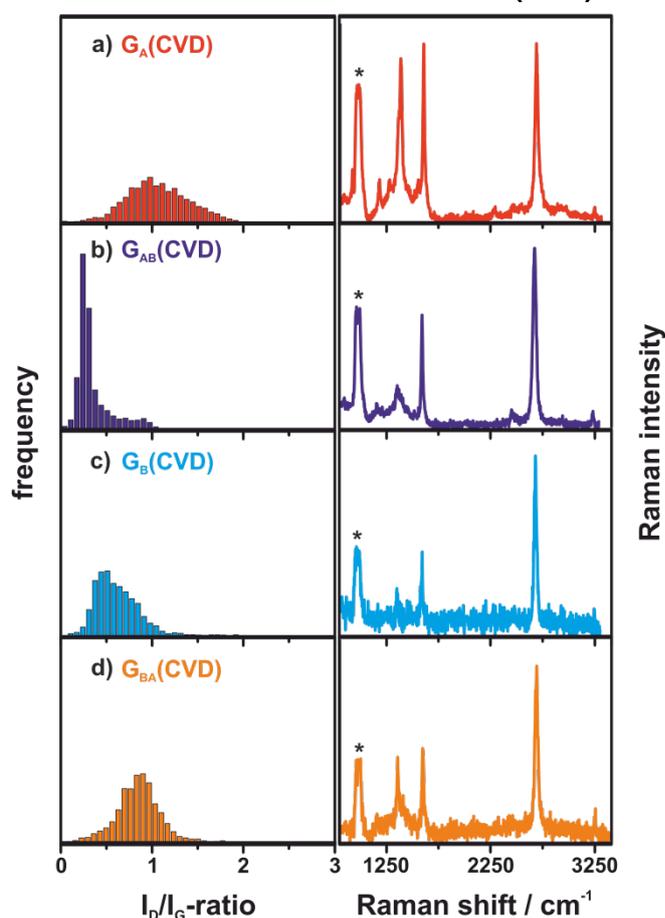

**Figure 4.** Left: Raman histograms (I$_D$/I$_G$-ratio) of **G$_A$(CVD)**, **G$_{AB}$(CVD)**, **G$_B$(CVD)**, and **G$_{BA}$(CVD)** on a Si/SiO$_2$ substrate. Right: exemplary point spectra of the respective samples, $\lambda_{exc.}$ = 532 nm.

Obviously, in this case the retro-reaction depicted in Scheme 4a is the predominant process. The difference between the two reaction sequences displayed in Scheme 4 can be explained with the fact that the aryl anion is the better leaving group compared with the alkyl anion. On the other hand, the difference between the double-sided bulk functionalization (Scheme 2, 3) and the single-sided functionalization (Scheme 4) is due to the fact that in the former case strain-free addition geometries can be adopted, whereas in the latter case the increasing degree of addition leads to decreased thermodynamic stability of the adducts (Scheme 5). This eventually leads, when sufficiently good leaving groups can be generated, to preferred retro-reactions. As pointed out above, exactly the same retro-reactions have previously been observed in fullerene and carbon nanotube chemistry, where only monotopic, exohedral addend binding can take place.[19-20]

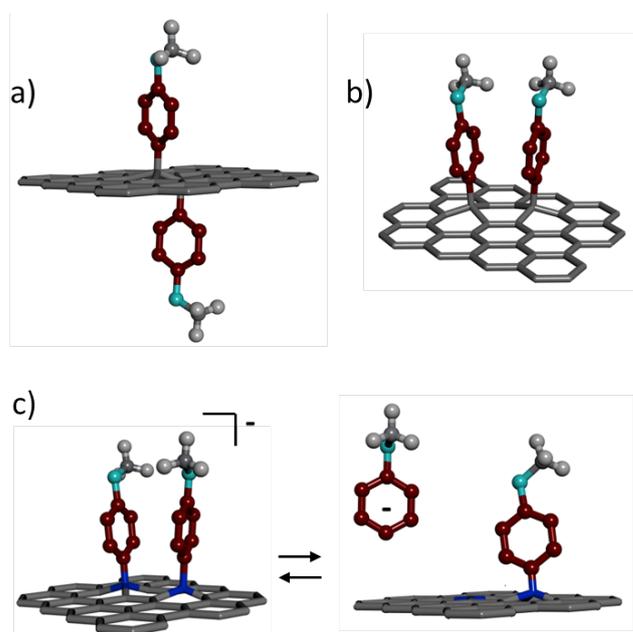

**Scheme 5.** Addition patterns of a) double-sided functionalized graphene (*trans*-1,2-addition) and b) single-sided functionalized graphene (*cis*-1,4-addition). Structure a) is more stable than b) due to the possibility of avoiding eclipsing 1,2-interactions and unfavourable bond angles. c) Equilibrium between retro-functionalization and multiple functionalization.

In conclusion, we have accomplished for the first time, bis-functionalization of graphene employing two successive reduction and covalent bond formation steps. Both bulk functionalization in solution and functionalization of individual sheets deposited on surfaces have been carried out. Whereas in the former case attacks from both sides of the basal plane are possible and can lead to strain-free architectures, in the latter case only monotopic attacks can be accomplished. As a consequence, a) strain energy due to eclipsing addend interactions and unfavourable geometries is increasingly built up and b) reduction induced retro-functionalization reaction can get important when the corresponding anion of the addend represents a sufficiently good leaving group. These fundamental reactivity studies of graphene chemistry will play an important role for the design of highly integrated graphene architectures with complex and tunable functionalities.

**Acknowledgements**

This work was carried out within the framework of the SFB 953 "Synthetic Carbon Allotropes". The research leading to these results has received partial funding from the European Union Seventh Framework Programme under grant agreement no.604391 Graphene Flagship. We would also like to thank the Graduate School Molecular Science for financial support.

# Supporting Information

# Double- and Single-Sided Multiple Functionalization of Graphene

Kathrin C. Knirsch,‡ Ricarda A. Schäfer,‡ Frank Hauke and Andreas Hirsch[*]

Department of Chemistry and Pharmacy & Institute of Advanced Materials and Processes
(ZMP), University of Erlangen-Nürnberg, Henkestr. 42, 91054 Erlangen, Germany
‡ The first two authors contributed equally to this work.

**Experimental Details:**

Instrumentation:

*Glove Box*

CVD graphene sample functionalization and preparation was carried out in an argon filled LABmaster[pro] sp glove box (MBraun), equipped with a gas purifier and solvent vapor removal unit: oxygen content < 0.1 ppm, water content < 0.1 ppm.

*Raman Spectroscopy*

Raman spectroscopic characterization was carried out on a LabRAM Aramis confocal Raman microscope (Horiba) with a laser spot size of about 1 µm (Olympus LMPlanFl 50x LWD, NA 0.50) in backscattering geometry. As excitation source a green laser with $\lambda_{exc}$ = 532 nm was used and the incident laser power was kept as low as possible (1.35 mW) to avoid any structural sample damage. Spectra were recorded with a CCD array at -70 °C – grating: 600 grooves/mm. Calibration in frequency was carried out with a HOPG crystal as reference.

*Thermogravimetric Analysis coupled with Mass Spectrometry (TGA-MS)*

Netzsch STA 409CD Skimmer with an EI ion source and a quadrupole mass spectrometer, using helium as inert gas; programmed time-dependent temperature profile: 24-100 °C (dynamic heating of 10 K/min), 100 °C for 30 min (isothermic heating step) and 100-800 °C (10 K/min temperature ramp).

*Infrared spectroscopy*

BRUKER Tensor® 27 FT-IR spectrometer with ATR sample holder.

Materials:

The flake graphite[1] *natural passau*† from Kropfmühl AG (Germany) was used as starting material. Chemicals and solvents were purchased from Sigma Aldrich and used without further treatment. Exempted from this is the solvent 1,2-dimethoxyethane (1,2-DME) which was several times distilled over Na/K alloy. Unlike the functionalization reagent hexyl iodide (Sigma Aldrich), the diazonium compound was synthesized (see below).

CVD graphene on a 0.5 x 0.5 cm$^2$ Si substrate with a 300 nm thick $SiO_2$ coating was purchased from Graphenea.

† pre-treatment of *natural passau* as starting material:

natural graphite flakes (Kropfmühl AG, Passau) are comminuted with 5 times the amount of sodium chloride in a mortar for 20 min, whereby, after the elution of NaCl with distilled water and drying in vacuum, smaller and easier dispersible graphite flakes are achieved than in the case of the original natural graphite flakes (see Supporting Information; K. C. Knirsch, J. M. Englert, C. Dotzer, F. Hauke and A. Hirsch, *Chem. Commun.*, 2013, **49**, 10811-10813.)

Experimental procedures:

*Synthesis of 4-methoxyphenyldiazonium tetrafluoroborate*

A solution of 2.12 g (17.2 mmol) 4-methoxyaniline in 50 mL THF was cooled to 0 °C. After the addition of 2.8 mL (3.56 g, 25.4 mmol) $BF_3$-THF and stirring for 5 min, 2.8 mL (2.44 g, 20.8 mmol) *iso*-pentylnitrite was added dropwise to the reaction solution at 0 °C which was then stirred for 15 min. The precipitated product was filtrated, washed with diethylether and dried under vacuum and simultaneous cooling with an ice bath. Yield: 3.50 g (15.8 mmol, 92 %)

**$^1$H-NMR** (400 MHz, DMSO-d6, RT): d = 8.60 (d, $^3J$ = 7.3 Hz, 2 H, *m*(OMe)-ArH), 7.47 (d, $^3J$ = 7.3 Hz, 2 H, *o*(OMe)-ArH), 4.03 (s, 3 H, $CH_3$) ppm.

**$^{13}$C{$^1$H}-NMR** (100 MHz, DMSO-d6, RT): d = 168.76 (1 C, ArC-OMe), 136.10 (2 C, *m*(OMe)-ArC-H), 117.24 (2 C, *o*(OMe)-ArC-H), 103.32 (1 C, ArC-$N_2^+$), 57.43 (1 C, $CH_3$) ppm.

**FT-IR** (ATR-solid): ñ = 3119 (Ar-H), 2265 (N$_2^+$), 1583 (aromatic ring), 1052 (B-F), 842 (1,4-disubstituted arene) cm$^{-1}$.

**EI** (+): *m/z* = 135 [M$^+$ - BF$_4^-$], 108 [M$^+$ - BF$_4^-$ - N$_2$].

*Bulk functionalization approach*

Under argon atmosphere in a Schlenk line, 24 mg (2.0 mmol carbon) graphite was added to the deep-blue solution of 0.45 mL freshly prepared, liquid Na/K alloy (eutectic mixture) in 150 mL dry, freshly distilled (over Na/K) 1,2-DME and stirred for 3 d. After short ultrasonic treatment and addition of the functionalization reagent (**A** or **B**, see Table S1), the reaction mixture was stirred overnight and worked-up with 50 mL cyclohexane and 50 mL distilled water. Washing with distilled water for three times (overall 200 mL), filtration through a 0.2 mm reinforced cellulose membrane filter (Sartorius) and final washing with isopropanol and distilled water (50 mL each), yielded a solid as product which was dried under vacuum overnight.

The resulting functionalized carbon material was used as starting material instead of pristine graphite for the second functionalization (same procedure as for the first functionalization) with the respective functionalization reagent of the other type (**B** or **A**, see Table S1), yielding the mixed functionalized material.

Table S1: Used amount of functionalization reagent.

| functionalization reagent | volume [mL] | mass [g] | amount of substance [mmol] | equivalents |
| --- | --- | --- | --- | --- |
| A | - | 2.66 | 12.00 | 6 |
| B | 3.00 | 4.24 | 20.00 | 10 |

**A**: 4-methoxyphenyldiazonium tetrafluoroborate

**B**: *n*-hexyl iodide

*Functionalization of CVD graphene*

In an argon filled glove box 130 mg Na/K were dissolved in 5 mL of DME and stirred for 1h. The diazonium salt MPD/hexyl iodide was dissolved in DME (0.1 mmol/mL). Two drops of the deep blue Na/K solution were used to activate the CVD graphene. Afterwards two drops of the electrophile dissolved in DME were added. The reaction was aborted after 10 min by removing reactants with DME. Outside of the glove box the wafer was washed with isopropanol, water and acetone. The resulting functionalized CVD graphene was used as starting material for the second addition which proceeded according to the first one.

Additional analytical data:

Figure S1: TG profile of the 4-methoxyphenyl-hexyl-functionalized reaction product **G**$_{AB}$ (black); Top: mass fragments of the 4-methoxyphenyl-addend. Center: m/z 39 shown as main mass fragment in addition to the other mass fragments of the 4-methoxyphenyl-addend. Bottom: mass fragments of the hexyl-addend.

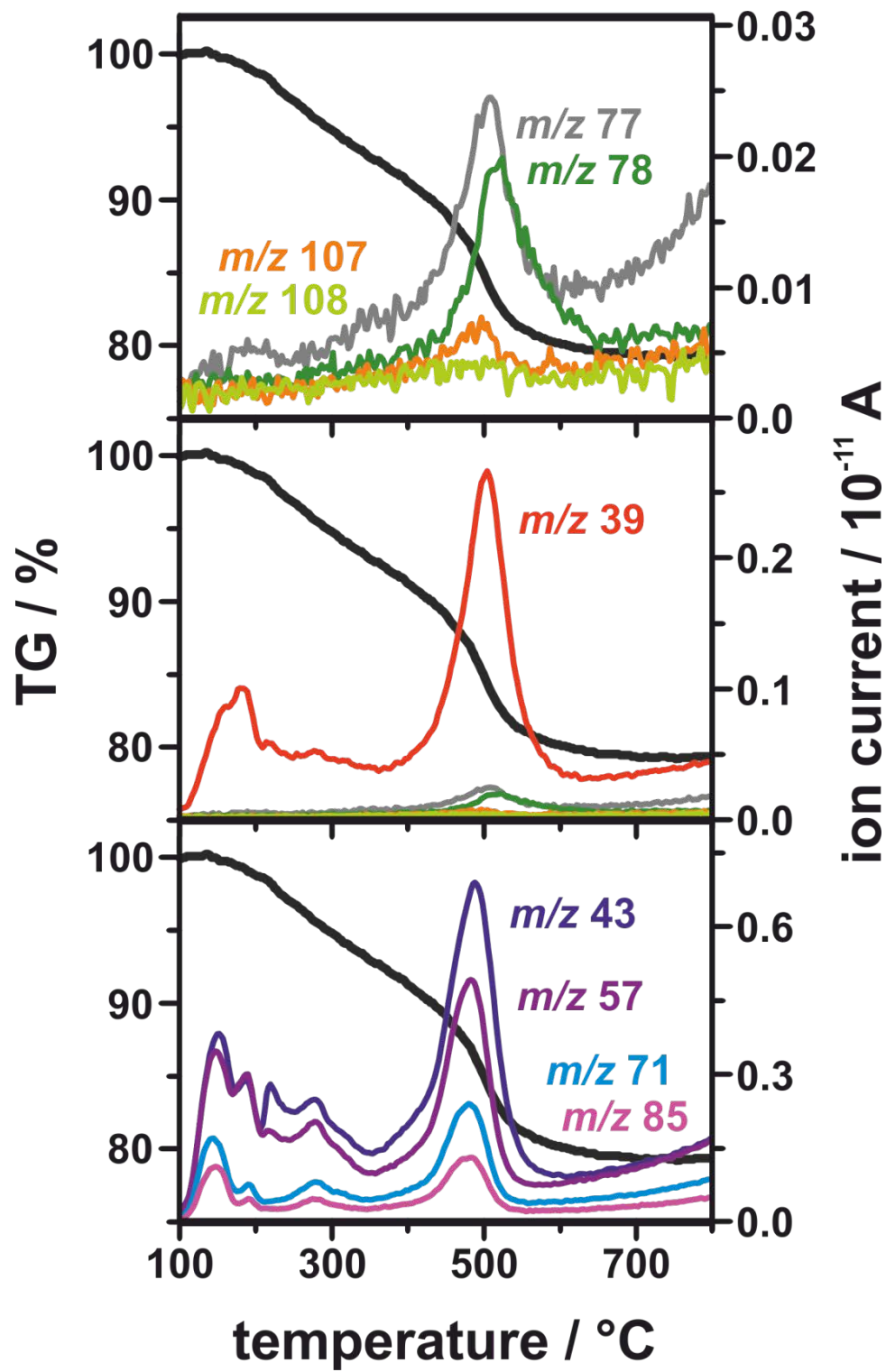

Scheme S1: MS-fragments of the 4-methoxyphenyl-hexyl-functionalized reaction product **G$_{AB}$** (TGA-MS measurement see Figure S1).

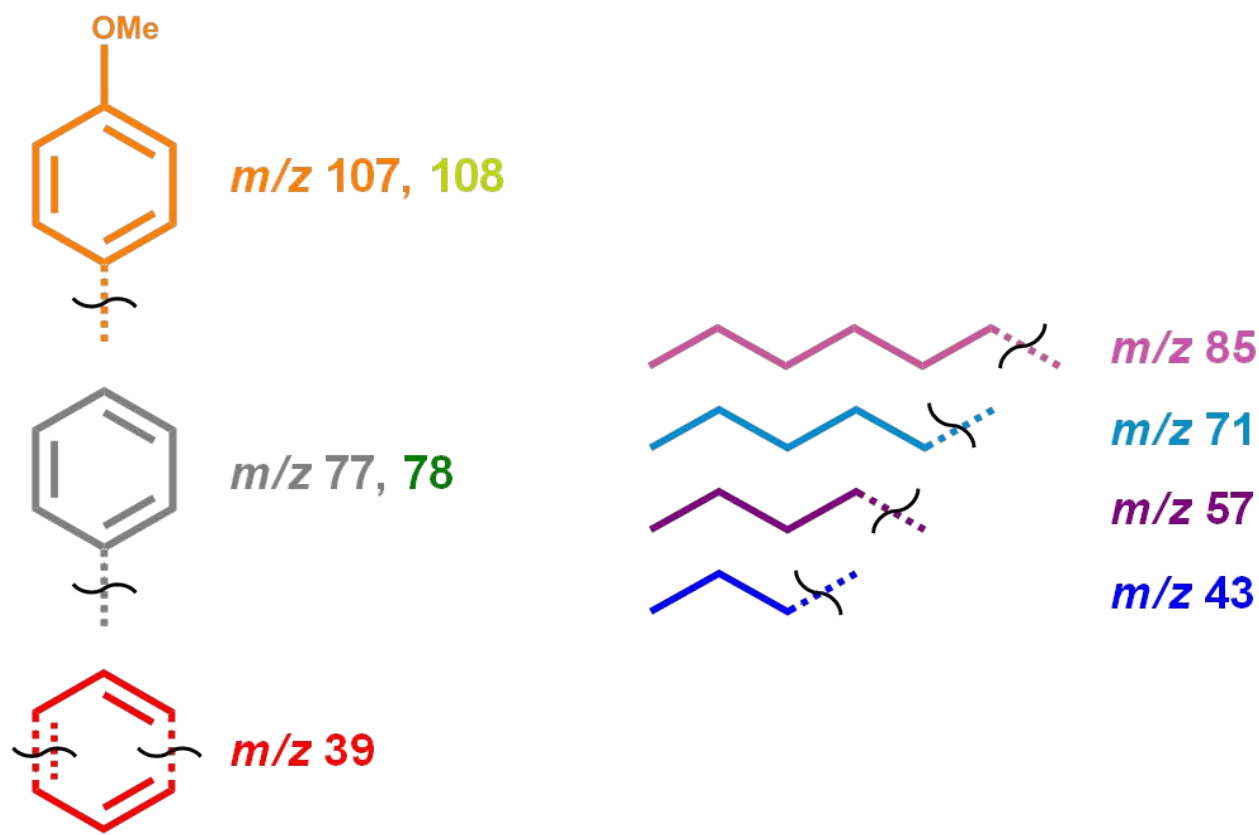

Figure S2: Left: Raman histogram (I$_D$/I$_G$-ratio) of **G$_P$**. Right: averaged spectrum of the sample, l$_{exc.}$ = 532 nm.

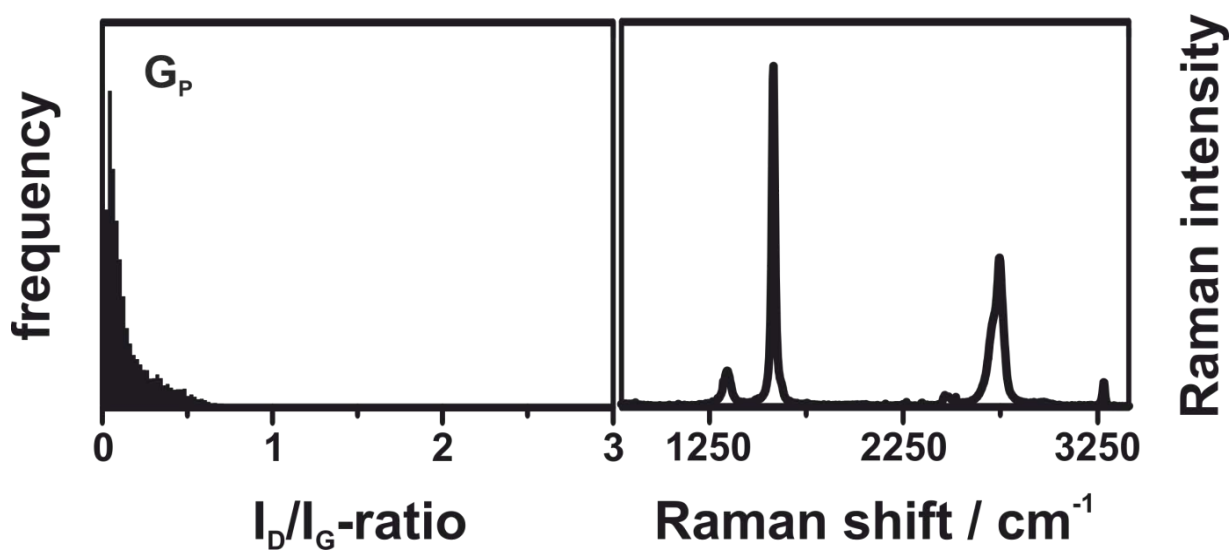

Table S2: Raman-data of **G$_P$** and the functionalized reaction products, $l_{exc.}$ = 532 nm.

| Sample | Raman ($I_D/I_G$)$_{average}$ | Sample | Raman ($I_D/I_G$)$_{average}$ |
|---|---|---|---|
| **G$_P$** | 0.1 | **G$_P$** | 0.1 |
| **G$_A$** | 0.7 | **G$_B$** | 1.6 |
| **G$_{AB}$** | 1.4 | **G$_{BA}$** | 1.3 |

Figure S3: Left: Raman histogram ($I_D/I_G$-ratio) of **G$_P$(CVD)**. Right: averaged spectrum of the sample, $\lambda_{exc.}$ = 532 nm (*: Si-O stretching vibration of the Si/SiO$_2$ substrate).

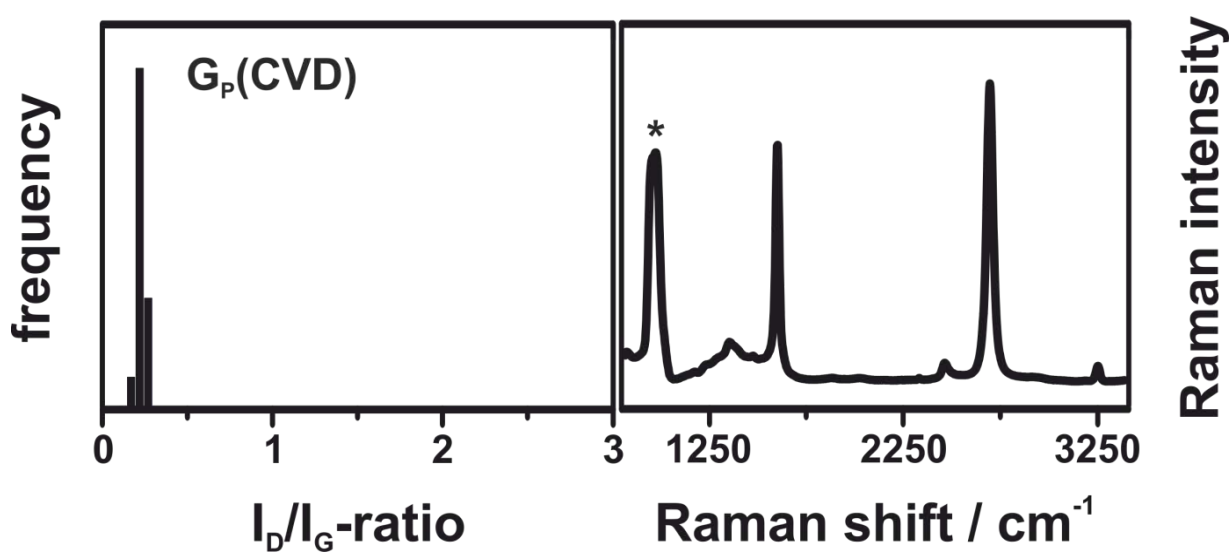

Figure S4: Raman map ($I_D/I_G$-ratio) of **G$_P$(CVD)**, $l_{exc.}$ = 532 nm.

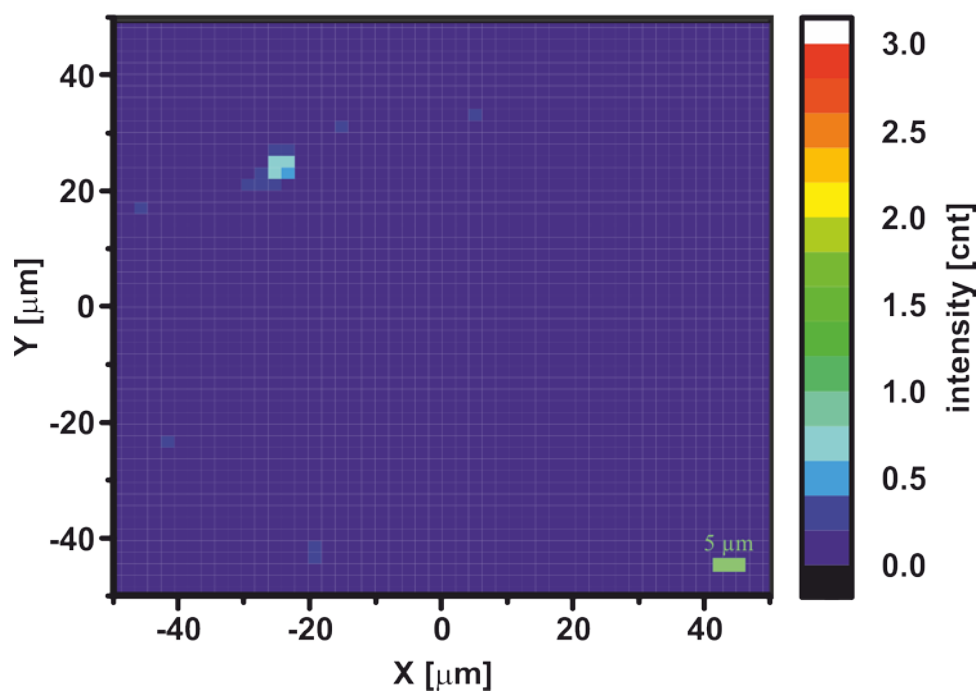

Figure S5: Left: Raman histogram (I$_{2D}$/I$_G$-ratio) of **G$_P$(CVD)**. Right: Raman histogram (FWHM$_{2D}$) of **G$_P$(CVD)**, $\lambda_{exc.}$ = 532 nm.

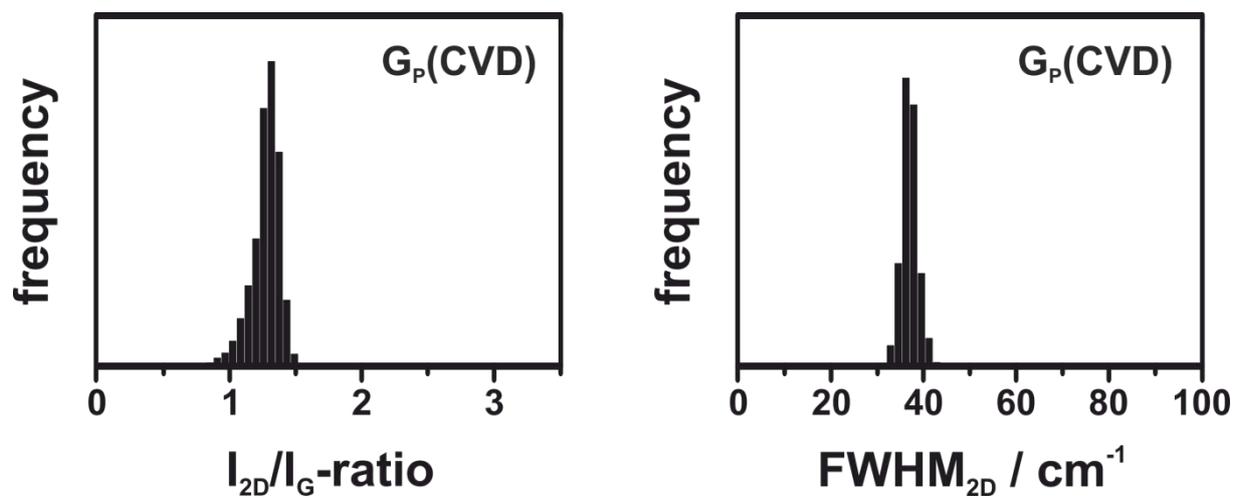

Figure S6: Raman map (I$_{2D}$/I$_G$-ratio) of **G$_P$(CVD)**, $\lambda_{exc.}$ = 532 nm.

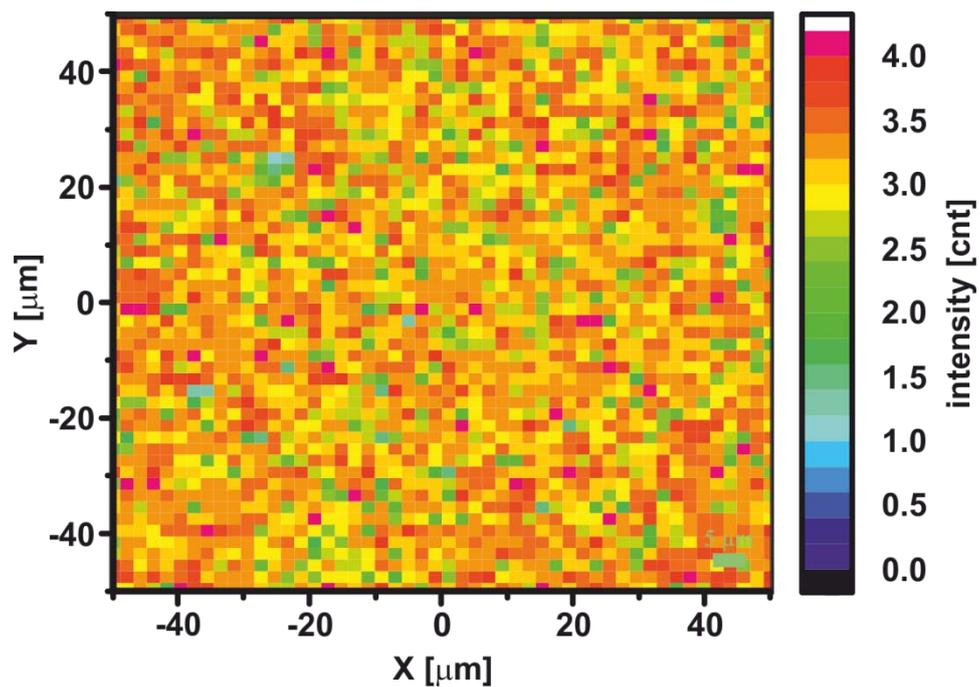

Figure S7: Raman map (FWHM$_{2D}$) of **G$_P$(CVD)**, $\lambda_{exc.}$ = 532 nm.

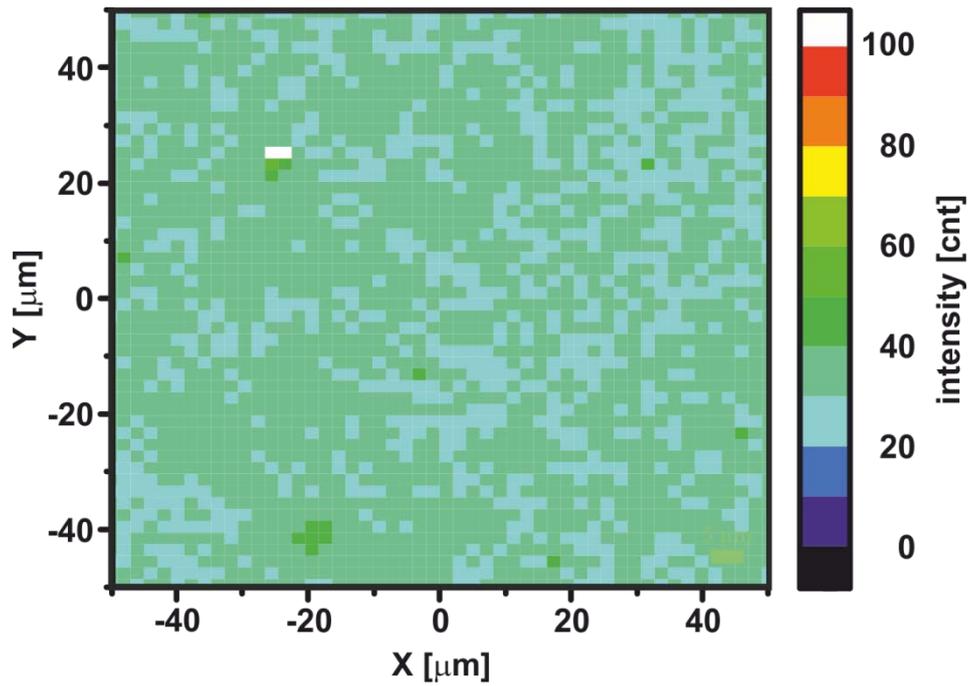

Figure S8: Left: Raman histogram (I$_D$/I$_G$-ratio) of **G$_P$(CVD) graphenide**. Right: averaged spectrum of the sample, $\lambda_{exc.}$ = 532 nm (*: Si-O stretching vibration of the Si/SiO$_2$ substrate).

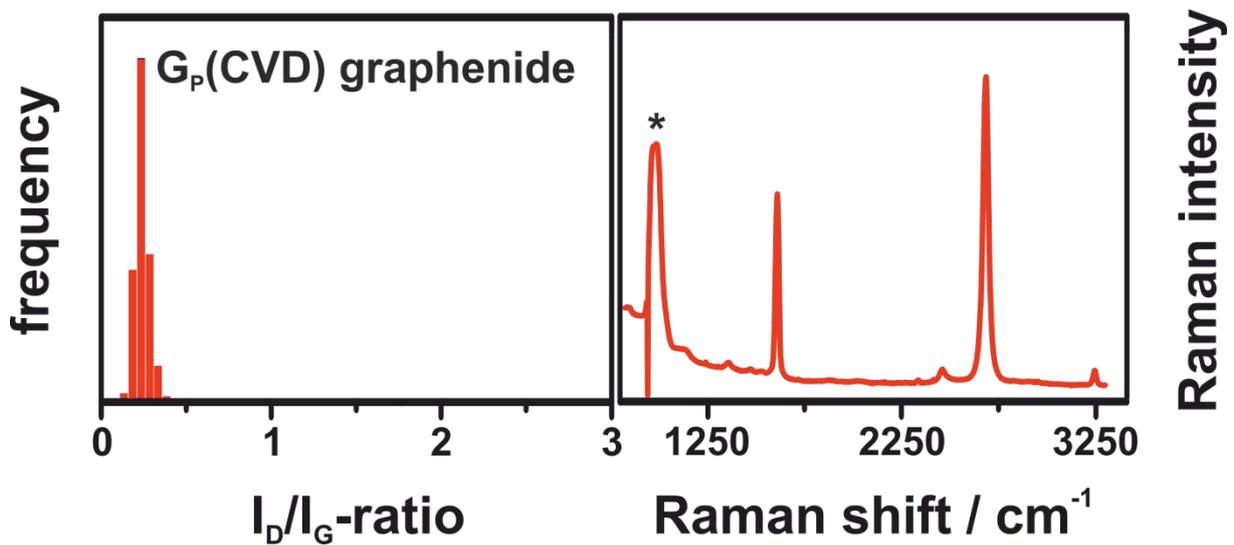

Figure S9: Left: Raman histogram ($I_{2D}/I_G$-ratio) of **$G_A$(CVD)**. Right: Raman histogram (FWHM$_{2D}$) of **$G_A$(CVD)**, $\lambda_{exc.}$ = 532 nm.

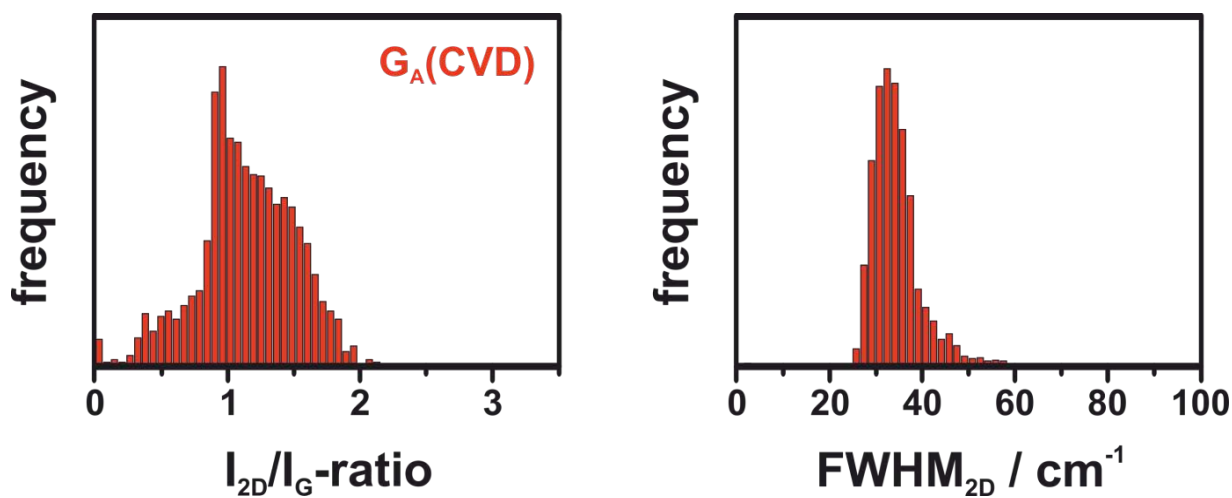

Figure S10: Left: Raman histogram ($I_{2D}/I_G$-ratio) of **$G_{AB}$(CVD)**. Right: Raman histogram (FWHM$_{2D}$) of **$G_{AB}$(CVD)**, $\lambda_{exc.}$ = 532 nm.

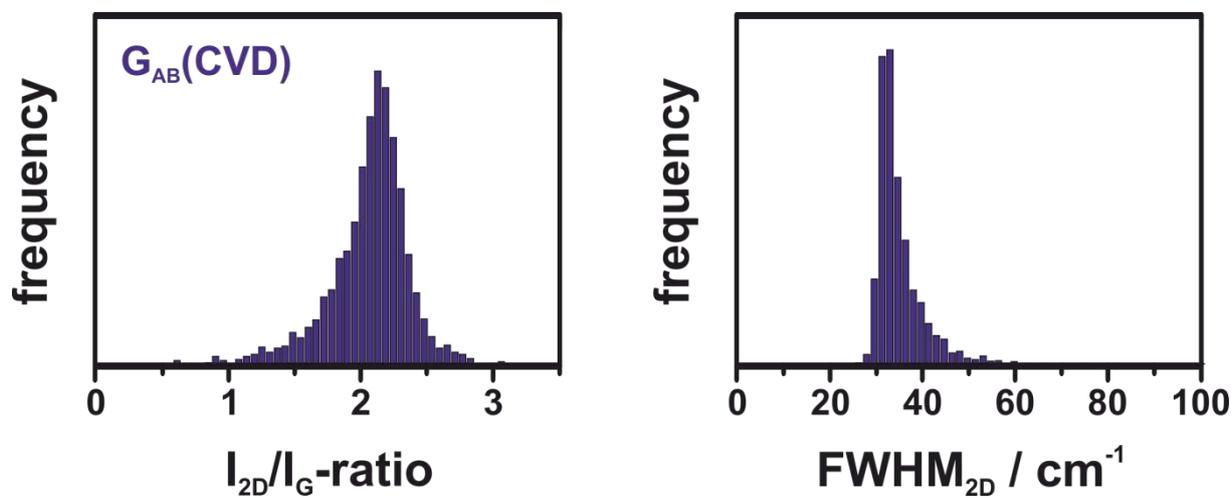

Figure S11: Left: Raman histogram ($I_{2D}/I_G$-ratio) of **G$_B$(CVD)**. Right: Raman histogram (FWHM$_{2D}$) of **G$_B$(CVD)**, $\lambda_{exc.}$ = 532 nm.

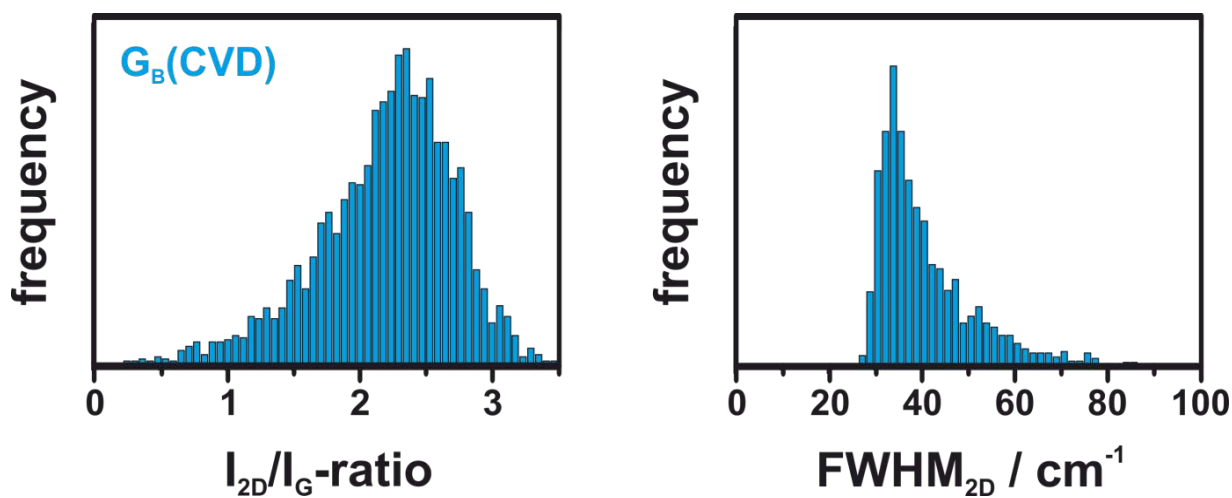

Figure S12: Left: Raman histogram ($I_{2D}/I_G$-ratio) of **G$_{BA}$(CVD)**. Right: Raman histogram (FWHM$_{2D}$) of **G$_{BA}$(CVD)**, $\lambda_{exc.}$ = 532 nm.

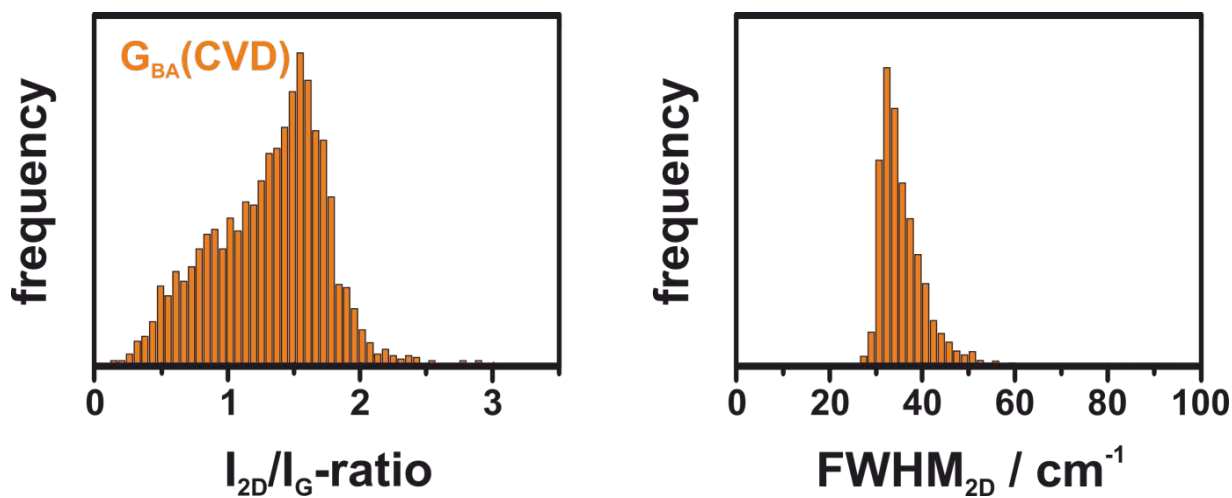

Table S3: Raman ($I_D/I_G$)$_{average}$ of **G$_P$(CVD)** and the functionalized reaction products, $\lambda_{exc.}$ = 532 nm.

| Sample | Raman ($I_D/I_G$)$_{average}$ | Sample | Raman ($I_D/I_G$)$_{average}$ |
|---|---|---|---|
| **G$_P$(CVD)** | 0.3 | **G$_P$(CVD)** | 0.3 |
| **G$_A$(CVD)** | 1.0 | **G$_B$(CVD)** | 0.6 |
| **G$_{AB}$(CVD)** | 0.3 | **G$_{BA}$(CVD)** | 0.9 |

Table S4: Raman $(I_{2D}/I_G)_{average}$ and $FWHM_{2D}$ of **G$_P$(CVD)** and the functionalized reaction products, $\lambda_{exc.}$ = 532 nm.

| Sample | Raman $(I_{2D}/I_G)_{average}$ | Raman $FWHM_{2D}$ |
|---|---|---|
| **G$_P$(CVD)** | 1.3 | 37 |
| **G$_A$(CVD)** | 1.0 | 33 |
| **G$_{AB}$(CVD)** | 2.1 | 32 |
| **G$_B$(CVD)** | 2.4 | 36 |
| **G$_{BA}$(CVD)** | 1.5 | 34 |